\documentstyle[prl,aps,floats,epsf,psfig,colordvi]{revtex}
\begin{document}
\draft
\newcommand{\rms}{\rm\scriptstyle}
\newcommand{\nub}{\overline{\nu}}
\newcommand{\Kethree}{\mbox{$K^{\pm}_{e3}$}}
\newcommand{\ecal}{\mbox{$E_{\rms cal}$}}
\wideabs{
\title{  A Search for $\nu_\mu \rightarrow \nu_e$
 and $\overline\nu_\mu \rightarrow  \overline\nu_e$
 Oscillations at NuTeV}
\author{
   S.~Avvakumov$^{8}$,  
 T.~Adams$^{4}$, A.~Alton$^{4}$,
 L.~de~Barbaro$^{5}$, P.~de~Barbaro$^{8}$, R.~H.~Bernstein$^{3}$, 
 A.~Bodek$^{8}$, T.~Bolton$^{4}$, J.~Brau$^{6}$, D.~Buchholz$^{5}$, 
 H.~Budd$^{8}$, L.~Bugel$^{3}$, J.~Conrad$^{2}$, R.~B.~Drucker$^{6}$, 
 B.~T.~Fleming$^{2}$, R.~Frey$^{6}$, J.A.~Formaggio$^{2}$, J.~Goldman$^{4}$, 
 M.~Goncharov$^{4}$, D.~A.~Harris$^{8}$, R.~A.~Johnson$^{1}$, J.~H.~Kim$^{2}$,
 S.~Koutsoliotas$^{2}$, M.~J.~Lamm$^{3}$, W.~Marsh$^{3}$, D.~Mason$^{6}$, 
 J.~McDonald$^{7}$, K.~S.~McFarland$^{8,3}$, C.~McNulty$^{2}$, 
    D.~Naples$^{7}$, 
 P.~Nienaber$^{3}$, V.~Radescu$^{7}$,  A.~Romosan$^{2}$, W.~K.~Sakumoto$^{8}$, H.~Schellman$^{5}$,
 M.~H.~Shaevitz$^{2}$, P.~Spentzouris$^{2}$, E.~G.~Stern$^{2}$, 
 N.~Suwonjandee$^{1}$, M.~Tzanov$^{7}$, M.~Vakili$^{1}$, A.~Vaitaitis$^{2}$, 
 U.~K.~Yang$^{8}$, J.~Yu$^{3}$, G.~P.~Zeller$^{5}$, and E.~D.~Zimmerman$^{2}$
}
\address{
$^1$University of Cincinnati, Cincinnati, OH 45221 \\
$^2$Columbia University, New York, NY 10027 \\
$^3$Fermi National Accelerator Laboratory, Batavia, IL 60510 \\
$^4$Kansas State University, Manhattan, KS 66506 \\
$^5$Northwestern University, Evanston, IL 60208 \\
$^6$University of Oregon, Eugene, OR 97403 \\
$^7$University of Pittsburgh, Pittsburgh, PA 15260 \\
$^8$University of Rochester, Rochester, NY 14627 \\ 
}
\date{\today}
\maketitle
\begin{abstract}
Limits on $\nu_\mu \rightarrow \nu_e$
 and $\overline\nu_\mu \rightarrow  \overline\nu_e$
oscillations   are 
  extracted using the NuTeV detector with sign-selected
   $\nu_\mu$ and $\nub_\mu$ beams.  In $\nub_\mu$ mode,
   for the case of $\sin^2 2\alpha = 1$, $\Delta m^2 > 2.6 $~${\rm eV^2}$
is excluded, and for
$\Delta m^2 \gg 1000$~${\rm eV^2}$, $\sin^2 2\alpha > 1.1 \times
10^{-3}$. The
  NuTeV data exclude  the high $\Delta m^2$  
   end of $\overline\nu_\mu \rightarrow  \overline\nu_e$
oscillations parameters favored by the LSND experiment
 without the need 
to assume that the oscillation parameters for $\nu$ and 
 $\nub$ are the same.
We present the most stringent experimental limits for
$\nu_\mu (\overline{\nu}_\mu) \to \nu_e (\overline{\nu}_e)$
oscillations in the large $\Delta m^2$
region. \\

\end{abstract}
\pacs{PACS numbers: 14.60.Pq, 13.15.+g \underline{\it ; UR-1640 to be 
published in Phys. Rev. Lett.}}
\twocolumn
}

Experimental  evidence for 
oscillations among the three neutrino generations has been recently reported.
For two-generation mixing, the
probability that a neutrino created as type $\nu_1$
oscillates to type $\nu_2$ is:
\begin{equation}
P(\nu_1 \rightarrow \nu_2) = \sin^2 2\alpha \sin^2 \left(\frac{1.27
\Delta m^2 L}{E_\nu}\right),
\label{eq:posc}
\end{equation}
where $\Delta m^2$ is the mass squared difference between the mass
eigenstates in ${\rm eV^2}$, $\alpha$ is the mixing angle, $E_\nu$ is
the incoming neutrino energy in $\rm GeV$, and $L$ is the distance between
the points of creation and detection in km.

Data from the Super-Kamiokande atmospheric neutrino experiment~\cite{ATM} have
been interpreted  as evidence for $\nu_\mu \rightarrow \nu_{\tau}$ oscillations
with  $\sin^2 2\alpha > 0.88$  and 
$1.6  \times 10^{-3} < \Delta m^2 < 4  \times 10^{-3}$
~${\rm eV^2}$.
The LSND experiment has reported~\cite{lsnd} a signal consistent with
$\bar{\nu}_\mu \rightarrow \bar{\nu}_e$ 
oscillations with $\sin^2 2\alpha \approx 10^{-2}$ and $\Delta m^2
\stackrel{>}{\scriptstyle\sim} 1$~$\rm eV{^2}$ .
 The  solar neutrino experiments, and most recently SNO~\cite{SNO} have
 reported evidence for oscillations of
 $\nu_e \rightarrow (\nu_\mu, \nu_{\tau})$ with
 $\Delta m^2 <10^{-3}{\rm eV^2}$. Within a three-generation mixing 
 scenario and  under the
 assumption that the $\Delta m^2$ values for $\nu$ and 
 $\nub$ are the same,
 it is not possible to simultaneously
 accommodate  
 the Super-Kamiokande, LSND,  and SNO results. Therefore,
 experimental searches for oscillations with both $\nu$ and  $\nub$ 
 beams are of interest. In this letter, we report on a search for 
 oscillations in both the
 $\nu_\mu \rightarrow \nu_e$
 and $\overline\nu_\mu \rightarrow  \overline\nu_e$ channels using
 a new sign-selected neutrino beam.

 High-purity $\nu$ and $\nub$ beams are provided by the new Sign-Selected
Quadrupole Train (SSQT) beamline at the Fermilab Tevatron during the 1996-1997
fixed target run.  Hadrons are produced when
the $800$~GeV primary proton beam interacts
in a BeO target located 1436 m upstream of the
neutrino detector. Sign-selected secondary particles of specified
charge (mean
energy of about 250 GeV) are directed in a 221 m beamline
towards a 320 m decay region, while oppositely
charged (and neutral) mesons are stopped in beam dumps.
Two-body decays of the focused pions
yield $\nu_\mu$  ($\overline\nu_\mu$) with a mean
energy of $\approx$75 GeV. 
Two-body decays of the focused kaons
yield $\nu_\mu$ ($\overline\nu_\mu$) with a mean
energy of $\approx$200 GeV. Muons are stopped
in a 915 m steel/earth shield.

\begin{figure} [bh]
\centerline{\psfig{figure=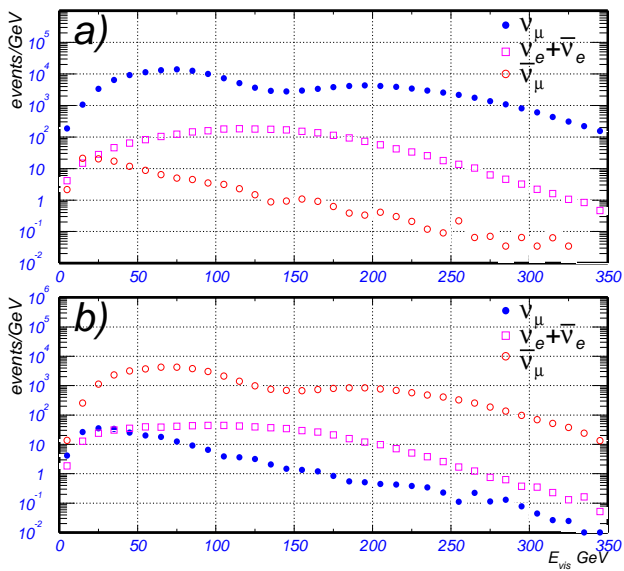,width=3.3in,height=3.0in}}
\caption{(a) The predicted visible energy spectra for 
 $\nu_\mu$, $\overline\nu_\mu$,  and 
($\nu_e$,$\overline\nu_e$) CC-events in $\nu_\mu$ running mode.
(b) The corresponding spectra for $\overline\nu_\mu$  running mode.
The predictions come from a beam
simulation tuned to agree with the observed number
of $\nu_\mu$ or $\overline\nu_\mu$  CC events in each 
running mode.}
\label{fig:enu}
\end{figure}

The energy
and spatial distributions
of $\nu_\mu$  ($\overline\nu_\mu$) CC events in the detector provide
a determination of the flux of pions and kaons in the decay channel
(used in the determination of the predicted $\nu_e$ and $\nub_e$ fluxes).
For $\nu_\mu$  running mode, 
the predicted energy spectra for 
 $\nu_\mu$, $\overline\nu_\mu$, and 
($\nu_e$+$\overline\nu_e$) CC events are shown in Figure \ref{fig:enu}(a).
The corresponding spectra for $\overline\nu_\mu$  running mode
are shown in Figure \ref{fig:enu}(b).
The  $\nu_\mu$ ($\overline\nu_\mu$)beam contains a 1.7\% (1.6\%)  
 $\nu_e$'s($\nub_e$'s) $93\%$ and 
$70\%$ of which are produced from $K^\pm \rightarrow \pi^0 e^\pm
\stackrel{_{(-)}}{\nu_e}$, for $\nu$ and $\overline\nu$ modes, 
respectively.
The proton beam is incident on the
production target at an angle such that forward neutral
kaons do not point at the detector.
This greatly reduces the electron neutrino flux from neutral kaon decays (which
is more difficult to model). The error in the predicted
electron neutrino flux is reduced from $4.1 \%$ (in CCFR~\cite{ALEX}) to $1.4 \%$ 
(NuTeV).

The NuTeV detector~\cite{NIM}
 is an upgrade of the CCFR detector ~\cite{CCFR}.
 It consists of an 18~m long, 690~ton
total absorption target-calorimeter with a mean density of ${\rm 4.2
~g/cm^3}$. Muon energy is measured by a 10~m long iron toroidal spectrometer. The
target consists of 168 steel plates, each ${\rm 3~m \times 3~m \times
5.15~cm}$, instrumented with liquid scintillation counters placed
every two steel plates and drift chambers 
every four plates.
The separation between consecutive scintillation counters corresponds to six
radiation lengths. 
The energy resolution of the target calorimeter is 
$\Delta E_h/E_h \approx 0.85/\sqrt{E_h}$(GeV), and $\Delta E_e/E_e \approx 
0.50/\sqrt{E_e}$(GeV)
for hadrons and electrons, respectively. The muon momentum resolution is
$\Delta p_\mu/p_\mu = 0.11$. 
The NuTeV detector
is calibrated continuously every accelerator
cycle (once a minute) 
with beams of electrons, muons,  and hadrons during
the slow spill part of the cycle.

While the neutrinos arrived in gates a few $msec$ wide, the
calibration beam arrived in a different gate 20 $sec$ long,
followed by an off-spill cosmic ray gate for background measurement.  
These continuous test beam calibrations yield a reduction in the hadron  
energy scale error from $1\%$ (in CCFR~\cite{CCFR}) to $0.43\% $ 
(in NuTeV~\cite{NIM}).

The event sample used in this analysis is similar to that
used in the recent precise NuTeV measurement
of the electroweak mixing angle~\cite{NCPRL} with additional
fiducial cuts, and $\ecal$ $>$ 30 GeV.
The data sample consists of 1.5 x 10$^{6}$ $\nu$
events
and 0.35 x 10$^{6}$ $\nub$ events with a mean visible energy
in the calorimeter ($\ecal$) of 74 
GeV and 56 GeV, respectively.
The observed neutrino events are separated
into CC and NC candidates. Both CC and
NC interactions initiate a cascade of hadrons in the target that
is registered in both the scintillation counters and drift chambers. 
Muon neutrino CC events are distinguished by the presence of a final
state muon, which typically penetrates well beyond the hadronic shower
and deposits energy in a large number of consecutive scintillation counters.
NC events usually have no final state muon and deposit energy over a
range of counters typical of a hadronic shower (about ten counters 
$\approx$ 1~m of steel).
For each event, 
the length ($L$) is defined as the number of scintillation counters between the
interaction vertex and the last counter consistent with at least single muon 
energy deposition.
 A pure sample of 
$\nu_{\mu}N \rightarrow \mu^-X$ $\nu_{\mu}$ charged current
events  is obtained from a $\lq$long' sample with 
$L \geq 29$ for $\nu$ running mode ($L \geq 28$ for $\nub$).
 The ``short'  event sample consists of events with
 $L \leq 28$ for $\nu$ running mode
($L \leq 27$ for $\nub$).
Events with a $\lq$short' length are primarily NC induced and 
originate from:
\begin{enumerate}
\item $\nu_{\mu,e}N \rightarrow \nu_{\mu,e}X$;  $\nu_{\mu,e}$ NC 
events;
($\approx 65\%$);
\item  $\nu_{\mu}N \rightarrow \mu^-X$;  $\nu_{\mu}$ short CC
events with muons which range out or exit the side of
the calorimeter ($25\%$ for $\nu$, $15\%$ for $\nub$);
\item $\nu_e N \rightarrow eX$ $\nu_e$ CC events
 ($10\%$ for $\nu$, $15\%$ for $\nub$); 
\item $\mu N \rightarrow \mu X$;  steep cosmic ray interactions 
($2\%$ for $\nu$ and $9\%$ for $\nub$).
\end{enumerate}

The electron produced in a $\nu_e$~CC event (source 3)
deposits energy in a few counters immediately downstream of the
interaction vertex;  this changes the longitudinal energy 
deposition profile of the
shower.
The  energy  profile is characterized by the ratio of the
sum of the energy deposited in the first three scintillation counters
to the total visible energy in the calorimeter $\ecal$:
\begin{equation}
\eta_3 \equiv \frac{E_1 + E_2 + E_3}{\ecal}, \label{eq:eta3}
\end{equation}
where $E_i$ is the energy deposited in the $i^{th}$ scintillation
counter downstream of the interaction vertex, and  $\ecal$ 
is the sum of the energy in the 20 scintillation counters downstream
(plus 1 upstream) of the vertex.
 We similarly define $\eta_2$ to be the ratio of the sum of the energy deposited
in the first two scintillation counters to $\ecal$. 

 Although the electromagnetic shower component is typically much shorter than a
hadron shower, NC and $\nu_e$~CC interactions cannot be separated 
on an event-by-event basis.
However, the difference in the shower profiles
can be used to perform a statistical extraction of the number of
$\nu_e$~CC events in the "$\lq$short' sample using the  $\eta_3$ distribution.

We first assume
that (for the same final state hadron energy) hadron showers produced in NC and
CC interactions are the same. Any difference in the 
$\eta_3$ (or   $\eta_2$) distributions
of $\lq$long' and $\lq$short' events is attributed to the
presence of $\nu_e$~CC interactions in the $\lq$short' sample. To compare
directly the $\lq$long' and $\lq$short' events,  a muon track from the data 
is added to the $\lq$short' events to compensate for the absence of a muon in
NC events. The fraction {\em f} of $\nu_\mu$~CC (source 2) events with a low
energy muon contained in the $\lq$short' sample which now have two muon
tracks is estimated from a detailed Monte Carlo of the experiment.
A simulated sample of such events is obtained by
choosing $\lq$long' events with the appropriate energy distribution from the
data to which a second short muon track is added in software. The
length of the short track and the angular distribution are obtained
from a Monte Carlo of $\nu_\mu$~CC events.

A sample of $\nu_e$ CC interactions in our detector is simulated by
adding a GEANT \cite{g321} generated electromagnetic shower of the
appropriate electron energy to 
the calorimetry counter energies 
in events in the $\lq$long' data sample. There is 
good  agreement between the GEANT simulation of electrons and the 
test beam data.
The energy distribution of the electron neutrinos and the fractional energy
transfer $y$ (in each event) are generated using a detailed Monte Carlo simulation of
the experiment. The CC cross section
model is tuned to agree with the measured CCFR differential cross
sections~\cite{yang} for $\nu_\mu$ events.
Since the hadron showers in the $\lq$long' sample already
have a muon track, this
sample ( $\nu_e CC +\mu$) can be compared directly with
the $\lq$short' and $\lq$long' samples.

The $\lq$long' and 
$\lq$short' $\eta_3$ distributions are further corrected by
subtracting the contamination due to cosmic ray events. 
 The cosmic ray component (source 4), which is only important for very
low energy bins, is well measured using the off-spill cosmic ray data.
Additionally, the $\eta_3$ distribution of short $\nu_\mu$~CC events 
(source 2),
normalized to the predicted fraction {\em f}, is subtracted from the
$\lq$short' event sample.

To extract the number of $\nu_e$~CC events in each $\ecal$
bin, we fit the corrected shape of the observed $\eta_3$ distribution
for the $\lq$short' sample to a combination of $\nu_\mu$~CC and $\nu_e$~CC
distributions with appropriate muon additions:
\begin{equation}
{\rm [short+ \mu] = \alpha \: [\nu_\mu CC] + \beta \: [\nu_e CC +
\mu]}.
\end{equation}

At this point we improve~\cite{AVVA}
over the previous CCFR
analysis~\cite{ALEX} by correcting for additional effects.
First, the hadron shower in CC events also
includes the contribution of photons that are radiated from the muon
during the CC scattering process. These photons are not present in
hadron showers of NC events.  A correction is
applied by using the PYTHIA Monte Carlo\cite{PYTHIA} 
to generate the spectrum of photons radiated by the muon in 
CC events. The parameters in PYTHIA that govern the emission of
photons are tuned to yield agreement with the radiative corrections 
formalism of deRujula \cite{RC}. 
Second, the procedures of adding a muon in software to
the  $\lq$short' sample (to model a $\lq$long' event) and of modeling electron
neutrino events by adding GEANT-generated electromagnetic showers to
hadron showers from $\lq$long' events are 
corrected for imperfect modeling  using a full 
LEPTO/GEANT/GHEISHA simulation
of the experiment and analysis procedures.  In the GEANT simulation of
neutrino events, the
Lund Model is used to generate the initial particle composition
of  hadron showers. 
The entire experimental procedure is simulated with the
beam Monte Carlo
($\nu_e$,$\nub_e$)  flux as input.  Modeling corrections
are extracted for each
$\ecal$ bin from the small difference between the
extracted ($\nu_e$,$\nub_e$) flux using
simulated data and the flux extracted using perfect
modeling in the Monte Carlo.

The absolute flux of $\nu_e$'s is taken as the average of
the results from analyses done using the $\eta_3$ and $\eta_2$
variables (the statistical error from
the $\eta_3$ analysis is used) . An additional systematic error (2.3\% in
neutrino mode and 0.6\% in antineutrino mode) is included to account for
this difference.
This systematic error for antineutrinos is smaller because the final
state positron carries a much larger fraction of the energy in $\nub_e$ CC 
events.
\begin{figure} [bh]
\centerline{\psfig{figure=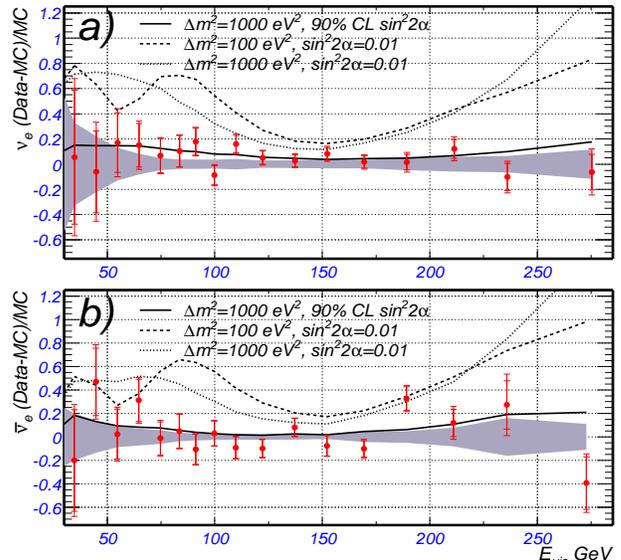,width=3.3in,height=3.0in}}
\caption{ The  ratio of
the detected over predicted numbers of
($\nu_e$,$\overline\nu_e$) events versus visible energy minus 1.
The curves correspond to the predictions
for oscillations  with $\sin^2 2\alpha =
0.01$, and  $\Delta m^2$ of $100$ and $1000$~${\rm eV^2}$.
The solid line is the 90\% confidence
upper limit for $\Delta m^2$=1000~${\rm eV^2}$.
The shaded area corresponds to the systematic error band.}
\label{fig:ratio}
\end{figure} 

For the oscillation search we measure the absolute flux of $\nu_e$'s
at the detector and compare it to the flux predicted by a detailed
beamline simulation \cite{NCPRL,AVVA}.
In order to extract limits on oscillations, 
the data are fitted by forming a $\chi^2$ which incorporates the Monte
Carlo generated effect of oscillations and terms
with coefficients accounting for systematic uncertainties. A best fit
$\sin^2 2\alpha$ is determined for each $\Delta m^2$ by minimizing
the $\chi^2$ as a function of $\sin^2 2\alpha$ and these systematic
coefficients.  
Figure \ref{fig:ratio} show the ratios of the measured rate of
($\nu_e$,$\nub_e$) CC events to
the Monte Carlo predictions minus 1. 
The inner
errors are the statistical errors. In order to show
the magnitude of the systematic errors, the outer errors include
all systematic errors added in quadrature (in the analysis
all correlations are taken into account). 
The shaded area  in the figure corresponds to the systematic error band
from uncertainties in the predicted electron flux
(primarily from the error in the  $K^\pm$ branching ratio)
 and uncertainties in the measured flux at the detector (primarily 
from the  $\eta_3$,$\eta_2$ difference, and the $2\%$ error in the 
electron energy scale). 
The curves correspond to the predictions
for oscillations  with $\sin^2 2\alpha =
0.01$, and  $\Delta m^2$ of  $100$ and $1000$~${\rm eV^2}$.
\begin{figure}[bh]
\centerline{\psfig{figure=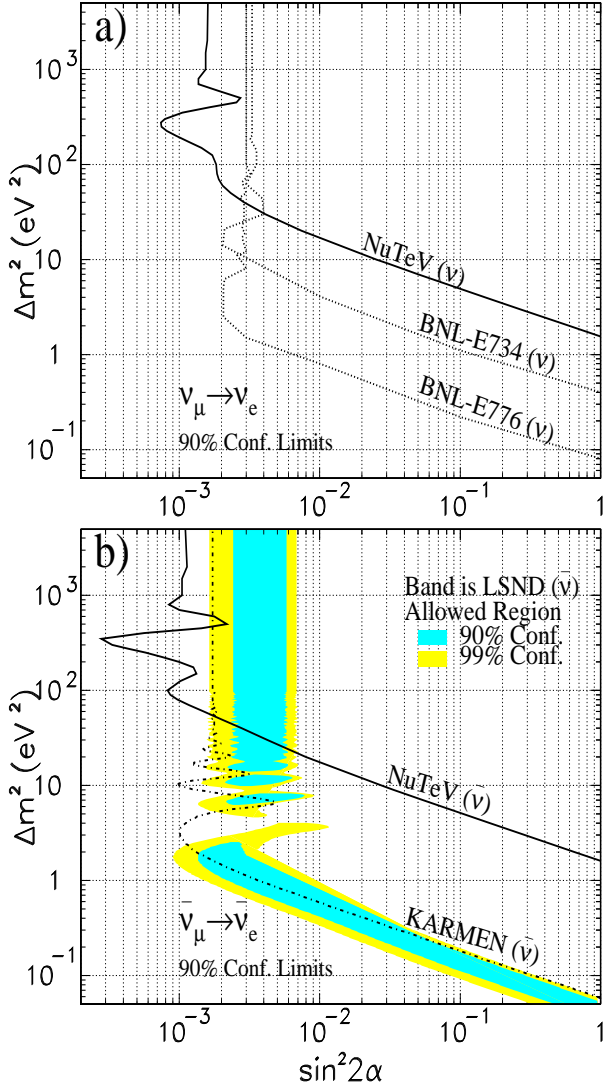,width=3.3in,height=5.7in}}
\caption{(a) Excluded region of $\sin^2 2\alpha$ and $\Delta m^2$ for
$\nu_\mu \rightarrow \nu_e$ oscillations from the NuTeV analysis at 90\%
confidence is the area to the right of the dark, solid curve. (b) NuTeV 
limits for $\overline\nu_\mu \rightarrow  \overline\nu_e$.
}
\label{fig:osc}
\end{figure}

At all $\Delta m^2$, the data are consistent with no
observed $\nu_\mu \rightarrow \nu_e$ oscillations (i.e. the best fit values
of $\sin^2 2\alpha$ 
are within one  standard deviation of zero). 
The frequentist approach \cite{pdg}
is used to set a 90\% confidence
upper limit for each $\Delta m^2$. The limit in $\sin^2 2\alpha$
corresponds to a shift of 1.64 units in $\chi^2$ from the minimum
(including all systematic errors).
The 90\% confidence upper limit is shown in Fig. \ref{fig:osc}(a) for
$\nu_\mu \rightarrow \nu_e$. Also shown are limits from
BNL-E734 \cite{e734} and 
BNL-E776 \cite{e776}. 
For $\sin^2 2\alpha = 1$, $\Delta m^2 > 2.4 $~${\rm eV^2}$ is excluded, and for
$\Delta m^2 \gg 1000$~${\rm eV^2}$, $\sin^2 2\alpha > 1.6 \times
10^{-3}$ (the best fit is at $1000$~${\rm eV^2}$ is  $\sin^2 2\alpha = 
(0.4 \pm 0.9 \times
10^{-3}$).
In the large $\Delta m^2$
region, NuTeV provides improved limits for  $\nu_\mu
\rightarrow \nu_e$ oscillations.

Similarly, the limit for  $\overline\nu_\mu \rightarrow  \overline\nu_e$
is shown Fig. \ref{fig:osc}(b). Also
shown are the LSND \cite{lsnd} results 
and limits from KARMEN ~\cite{karm}.
For the case of $\sin^2 2\alpha = 1$, $\Delta m^2 > 2.6 $~${\rm eV^2}$ is excluded, and for
$\Delta m^2 \gg 1000$~${\rm eV^2}$, $\sin^2 2\alpha > 1.1 \times
10^{-3}$ (the best fit is at $1000$~${\rm eV^2}$ is  $\sin^2 2\alpha = 
(-0.3 \pm 1.1 \times
10^{-3}$).  In  the $\nub_\mu$ mode, 
our results 
exclude  the high $\Delta m^2$ 
end of $\overline\nu_\mu \rightarrow  \overline\nu_e$
oscillations parameters favored by the LSND experiment,
without the need
to assume that the oscillation parameters for $\nu$ and 
 $\nub$ are the same.
  These 
are the most stringent experimental limits~\cite{karm} for
 $\nu_\mu (\overline{\nu}_\mu) \to \nu_e (\overline{\nu}_e)$
 oscillations in the large $\Delta m^2$ region.


This work was supported by the U.S. Department of Energy,
the National Science Foundation, and the Alfred P. Sloan
foundation.
\end{document}